\documentclass[twocolumn]{aastex62}
\usepackage{float}

\graphicspath{{./}{figures/}}

\shorttitle{A 20 minute binary}
\shortauthors{Burdge et al.}

\begin{document}

\title{Orbital Decay in a 20 Minute Orbital Period Detached Binary with a Hydrogen Poor Low Mass White Dwarf}

\correspondingauthor{Kevin B. Burdge}
\email{kburdge@caltech.edu}

\author[0000-0002-7226-836X]{Kevin B. Burdge}
\affiliation{Division of Physics, Mathematics and Astronomy, California Institute of Technology, Pasadena, CA 91125, USA}
\author[0000-0002-4544-0750]{Jim Fuller}
\affiliation{Division of Physics, Mathematics and Astronomy, California Institute of Technology, Pasadena, CA 91125, USA}
\author{E. Sterl Phinney}
\affiliation{Division of Physics, Mathematics and Astronomy, California Institute of Technology, Pasadena, CA 91125, USA}
\author[0000-0002-2626-2872]{Jan van Roestel}
\affiliation{Division of Physics, Mathematics and Astronomy, California Institute of Technology, Pasadena, CA 91125, USA}
\author[0000-0002-4045-8134]{Antonio Claret}
\affiliation{Instituto de Astrof\'{\i}sica de Andaluc\'{\i}a,
  CSIC, Apartado 3004, 18080 Granada, Spain \\ 
  Dept. F\'{\i}sica Te\'{o}rica y del Cosmos, Universidad de Granada, Campus de Fuentenueva s/n,  10871, Granada, Spain}
\author[0000-0002-3184-3428]{Elena Cukanovaite}
\affiliation{Department of Physics, University of Warwick, Coventry, CV4 7AL, UK}
\author[0000-0002-6428-4378]{Nicola Pietro Gentile Fusillo}
\affiliation{Department of Physics, University of Warwick, Coventry, CV4 7AL, UK}
\author[0000-0002-8262-2924]{Michael W. Coughlin}
\affiliation{Division of Physics, Mathematics and Astronomy, California Institute of Technology, Pasadena, CA 91125, USA}
\author[0000-0001-6295-2881]{David L. Kaplan}
\affiliation{Department of Physics, University of Wisconsin-Milwaukee, Milwaukee, WI, USA}
\author[0000-0002-6540-1484]{Thomas Kupfer}
\affiliation{Kavli Institute for Theoretical Physics, University of California Santa-Barbara, Santa Barbara, CA, USA}
\author[0000-0001-9873-0121]{Pier-Emmanuel Tremblay}
\affiliation{Department of Physics, University of Warwick, Coventry, CV4 7AL, UK}
\author{Richard G. Dekany}
\affiliation{Caltech Optical Observatories, California Institute of Technology, Pasadena, CA, USA}
\author[0000-0001-5060-8733]{Dmitry A. Duev}
\affiliation{Division of Physics, Mathematics and Astronomy, California Institute of Technology, Pasadena, CA 91125, USA}
\author{Michael Feeney}
\affiliation{Caltech Optical Observatories, California Institute of Technology, Pasadena, CA, USA}
\author[0000-0002-0387-370X]{Reed Riddle}
\affiliation{Caltech Optical Observatories, California Institute of Technology, Pasadena, CA, USA}
\author[0000-0001-5390-8563]{S. R. Kulkarni}
\affiliation{Division of Physics, Mathematics and Astronomy, California Institute of Technology, Pasadena, CA 91125, USA}
\author[0000-0002-8850-3627]{Thomas A. Prince}
\affiliation{Division of Physics, Mathematics and Astronomy, California Institute of Technology, Pasadena, CA 91125, USA}



\begin{abstract}

We report the discovery of a detached double white dwarf binary with an orbital period of $\approx20.6\,\rm minutes$, PTF J053332.05+020911.6. The visible object in this binary, PTF J0533+0209B, is a $\approx 0.17\,M_\odot$ mass white dwarf with a helium-dominated atmosphere containing traces of hydrogen (DBA). This object exhibits ellipsoidal variations due to tidal deformation, and is the visible component in a single-lined spectroscopic binary with a velocity semi-amplitude of $K_B=618.7\pm6.9 \, \rm km \, s^{-1}$. We have detected significant orbital decay due to the emission of gravitational radiation, and we expect that the \emph{Laser Interferometer Space Antenna} (\emph{LISA}) \citep{2017arXiv170200786A} will detect this system with a signal to noise of $8.4^{+4.2}_{-3.0}$ after four years of operation. Because this system already has a well determined orbital period, radial velocity semi-amplitude, temperature, atmospheric composition, surface gravity, and orbital decay rate, a \emph{LISA} signal will help fully constrain the properties of this system by providing a direct measurement of its inclination. Thus, this binary demonstrates the synergy between electromagnetic and gravitational radiation for constraining the physical properties of an astrophysical object.

\end{abstract}



\section{Introduction}

After expanding into red giants at the end of their lives, most stars leave behind dense white dwarf remnants. Stars in binary systems can engulf their companions during this process and leave behind compact binaries with orbital periods of hours to days, and when this process occurs twice, it can produce double white dwarf binaries with orbital periods less than an hour \citep{2013A&ARv..21...59I}. According to general relativity \citep{1916SPAW.......688E}, these binary systems emit gravitational radiation at twice their orbital frequency. Although few such gravitational wave sources are currently known \citep{2018MNRAS.480..302K}, the \emph{Laser Interferometer Space Antenna} (\emph{LISA}) should detect tens of thousands of systems within the galaxy \citep[e.g.][]{2012ApJ...758..131N}. This emission of gravitational radiation causes the orbits of these systems to decay, and can result in orbital periods as short as a few minutes.

The few known binary systems emitting gravitational radiation with sufficient strain for \emph{LISA} to detect have been referred to as \emph{LISA} ``verification" binaries, as they will serve as tests that the detector is operating as expected; however, these binaries are rich probes of astrophysics which provide insight into binary evolution, the population of Type Ia progenitors, white dwarf physics, tidal physics, accretion physics, and are not simply sources which will ``verify" \emph{LISA}'s functionality. Most of the known \emph{LISA} detectable binaries are mass transferring AM CVn systems \citep{2018MNRAS.480..302K}. These are strong \emph{LISA} gravitational wave sources due to their short orbital periods; however, constraining their physical parameters is complicated by the accretion they undergo, which dominates their optical luminosity and determines their period evolution, thereby preventing the measurement of a chirp mass due to decay induced by gravitational wave emission. Detached eclipsing double white dwarf binaries, such as ZTF J153932.16+502738.8 \citep{Burdge2019} and SDSS J065133.338+284423.37 \citep{2011ApJ...737L..23B} are the best characterized \emph{LISA} gravitational-wave sources known, with precisely measured system parameters; however, currently only these two systems have been characterized with such precision.

We report the discovery of PTF J053332.05+020911.6 (hereafter referred to as PTF J0533+0209), a detached double white dwarf binary system with an orbital period of $1233.97298\pm0.00017 \,\rm s$. In this binary, there is only one visible component in the optical spectrum, and thus, this system is a single-lined spectroscopic binary. The system exhibits relativistic Doppler beaming, ellipsoidal modulation, and orbital decay due to the emission of gravitational radiation. We will refer to the unseen massive companion in this binary as PTF J0533+0209A, and the lower mass, tidally deformed, visible component of this binary as PTF J0533+0209B. In this paper, we report our measurements of the observable quantities of this system and present the physical parameters we infer from these (Table \ref{tab:Parameters}). We conclude by discussing the future study of detached, non-eclipsing systems like PTF J0533+0209 in the era of \emph{LISA}.

\section{Observations}

\begin{table}[h!]
\renewcommand{\thetable}{\arabic{table}}
\centering
\caption{Table of observed parameters} \label{tab:ObservableParms}
\begin{tabular}{ccD@{$\pm$}D}
\tablewidth{0pt}
\hline
\hline
Gaia& RA & $83.383588013\rm\,deg\pm0.34\,\rm mas$     \\
   \hline
   Gaia& Dec & $+2.153208645\rm\,deg\pm0.33\,\rm mas$      \\
   \hline
   Gaia& Parallax & $0.47\pm0.48\,\rm mas$     \\
   \hline
   Gaia& pm RA & $1.43\pm0.74\,\rm mas\,yr^{-1}$    \\
   \hline
   Gaia& pm Dec & $2.56\pm0.91\,\rm mas\,yr^{-1}$    \\
   \hline
   GALEX& NUV & $20.38\pm0.28\,m_{\rm AB}$     \\
   \hline
   Pan-STARRS & g & $19.00\pm0.02\,m_{\rm AB}$   \\
   \hline
   Pan-STARRS&r & $19.15\pm0.01\,m_{\rm AB}$   \\
   \hline
   Pan-STARRS&i & $19.40\pm0.01\,m_{\rm AB}$   \\
   \hline
   Pan-STARRS&z & $19.60\pm0.03\,m_{\rm AB}$   \\
  \hline
  Pan-STARRS&y & $19.61\pm0.06	\,m_{\rm AB}$   \\
  \hline

\end{tabular}
\end{table}

\subsection{Photometric Color Selection}
We discovered PTF J0533+0209 during a broad search for post common envelope binaries \citep{2013A&ARv..21...59I}. In order to target hot young remnants of the common envelope phase, we used the Pan-STARRS DR1 \citep{2016arXiv161205560C} photometric survey to target blue objects, selecting all objects with a color satisfying the condition $(g-r)<0$ (see Table \ref{tab:ObservableParms}--note that the values in the table are observed apparent magnitudes, and have not been de-reddened).

\subsection{PTF Photometry}

After imposing the photometric color cut, we cross-matched the resulting sample with the archival Palomar Transient Factory (PTF) \citep{2009PASP..121.1395L} photometric database. The Palomar Transient Factory (and the Intermediate Palomar Transient Factory) was a northern-sky synoptic survey using the 48-inch Samuel Oschin Telescope at Palomar Observatory. The survey was conducted in PTF $r$ and $g$ bands with a typical exposure time of $60\,\rm s$, resulting in limiting magnitudes of approximately $21$ and $20$, respectively. We restricted ourselves to a search in $r$ band because it is the most heavily sampled of the PTF bands. Additionally, we required a minimum of $20$ epochs in the lightcurves. This cross match yielded $\approx180,000$ sources. We discovered PTF J0533+0209 via a period search. We detected the object with high significance because it falls in the most heavily sampled field in all of PTF, located inside the constellation of Orion, which was observed over $5000$ times, primarily in the first two years of PTF. This field did not use the typical PTF exposure time of $60\,s$, but instead used $30\,s$ exposures.  These observations were taken in two brief intervals, one consisting of $\approx3000$ observations in Dec 2009-Jan 2010, and another in Dec 2010 with $\approx2000$ observations (Table \ref{tab:Observations}).

\subsection{High Speed Photometry}

\begin{figure}[htpb]
\begin{center}
\includegraphics[width=0.47\textwidth]{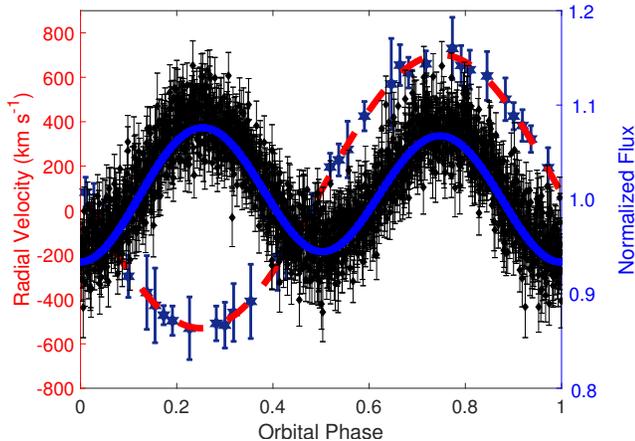}
\caption{Light curve model (solid blue line) overplotted with $g^\prime$ phase folded CHIMERA lightcurve, and radial velocity curve derived from LRIS spectra (dashed red line). The CHIMERA lightcurve exhibits ellipsoidal modulation, a geometric effect in which the brightness of the binary is modulated at twice the orbital frequency due to the tidal deformation of one object. The CHIMERA lightcurve also exhibits relativistic Doppler beaming manifested in the maximum at the phase of highest blueshift (Orbital Phase 0.25) being a few percent higher in relative flux than the phase of maximum redshift (Orbital Phase 0.75). A sinusoidal fit to the radial velocity curve yields a semi-amplitude of $K=618.7\pm6.9 \, \rm km \, s^{-1}$.}
\label{fig:KeckChimera}
\end{center}
\end{figure}

We used the high speed photometer on 200-inch Hale telescope at Palomar Observatory, CHIMERA \citep{2016MNRAS.457.3036H}, to obtain a well sampled follow-up lightcurve of the object. The instrument consists of a pair of electron multiplying charge-coupled devices (EM CCDs), has a dichroic and dual channels, allowing us to obtain simultaneous observations in $g^\prime$ and $i^\prime$. We used $10\, \rm s$ exposures for all observations operating with the conventional amplifier, and these lightcurves served as the basis for our analysis of the ellipsoidal modulation and relativistic Doppler beaming exhibited by the visible component in the system (Figure \ref{fig:KeckChimera}).

Additionally, we used the Kitt Peak Electron Multiplying CCD demonstrator (KPED) \citep{2019MNRAS.485.1412C}, a high speed EM CCD photometer mounted on Kitt Peak National Observatory's 84-inch telescope, to obtain additional observations to use as timing epochs (in order to measure the orbital decay). We obtained observations in $g^\prime$. Unlike our CHIMERA observations, we operated KPED with electron multiplying gain enabled, effectively eliminating read noise. The detector acquired images at a rate of 8 Hz, which we then stacked to $10\, \rm s$ coadditions in order to match the exposure time used with CHIMERA. Table \ref{tab:Observations} gives a summary of these observations. 

\begin{table*}
 \centering
 \caption{}
  \begin{tabular}{cllll}
  \hline\hline
   Instrument & Filter & Date & \# of Exposures & Exposure Time \\
        PTF & PTF $r$  &  Dec 4 2009-Jan 15 2010 & 3020  &  30s \\
     PTF & PTF $r$  &  Dec 9-15 2010 & 1775  &  30s \\
  \hline
     CHIMERA & $g^\prime$  &  Dec 14 2017 & 1600  &  10s \\
     CHIMERA & $i^\prime$  &  Dec 14 2017 & 1600  &  10s \\
  \hline
 CHIMERA & $g^\prime$  &  Dec 15 2017 & 1000  &  10s \\
 CHIMERA & $i^\prime$  &  Dec 15 2017 & 1000  &  10s \\
    \hline
  CHIMERA & $g^\prime$  &  Sept 17 2018 & 700  &  10s \\
 CHIMERA & $i^\prime$  &  Sept 17 2018 & 700  &  10s \\
   \hline
 CHIMERA & $g^\prime$  &  Sept 18 2018 & 700  &  10s \\
 CHIMERA & $i^\prime$  &  Sept 18 2018 & 700  &  10s \\
   \hline
 CHIMERA & $g^\prime$  &  Dec 31 2018 & 719  &  10s \\
 CHIMERA & $i^\prime$  &  Dec 31 2018 & 719  &  10s \\
  \hline
   KPED & $g^\prime$  &  Sept 8 2018 & 716 (stacked) &  10s (stacked)\\
  \hline
   KPED & $g^\prime$  &  Sept 9 2018 & 804 (stacked)  &  10s (stacked) \\
  \hline
   KPED & $g^\prime$  &  Sept 10 2018 & 737 (stacked)  & 10s (stacked) \\
  \hline
  KPED & $g^\prime$  &  Sept 11 2018 & 206 (stacked)  & 10s (stacked) \\
  \hline
  KPED & $g^\prime$  &  Sept 16 2018 & 707 (stacked)  & 10s (stacked) \\
  \hline
  KPED & $g^\prime$  &  Sept 17 2018 & 759 (stacked)  & 10s (stacked) \\
  \hline
  KPED & $g^\prime$  &  Dec 9 2018 & 1546 (stacked)  & 10s (stacked) \\
  \hline
  KPED & $g^\prime$  &  Dec 10 2018 & 236 (stacked)  & 10s (stacked) \\
  \hline
  KPED & $g^\prime$  &  Feb 26 2019 & 706 (stacked)  & 10s (stacked) \\
  \hline
  KPED & $g^\prime$  &  Mar 23 2019 & 668 (stacked)  & 10s (stacked) \\
  \hline
  KPED & $g^\prime$  &  Mar 29 2019 & 490 (stacked)  & 10s (stacked) \\
  \hline
  KPED & $g^\prime$  &  Mar 31 2019 & 344 (stacked)  & 10s (stacked) \\
  \hline
  KPED & $g^\prime$  &  Apr 2 2019 & 204 (stacked)  & 10s (stacked) \\
  \hline
  LRIS & Blue Arm  &  Nov 15 2017 & 18  & 120s \\
  \hline
  LRIS & Blue Arm  &  Mar 19 2018 & 40  & 120s \\
  \hline
\end{tabular}
\label{tab:Observations}
\end{table*}

All photometric data were reduced using a custom pipeline.

\subsection{Spectroscopy}

On October 25th, 2017, we observed the object on the Hale Telescope with four consecutive 5 minute exposures using the Double Spectrograph \citep{1982PASP...94..586O}. While these spectra have low signal to noise (SNR), they nonetheless reveal the presence of large Doppler shifts on the order of several hundred km~s$^{-1}$. We reduced this data using the pyraf-dbsp reduction pipeline \citep{2016ascl.soft02002B}.

\begin{figure}[htpb]
\begin{center}
\includegraphics[width=0.47\textwidth]{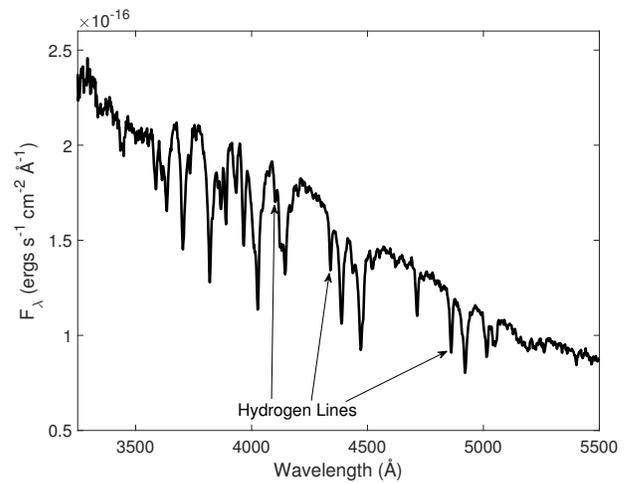}
\caption{The coadded LRIS spectrum of the PTF J0533+0209B. We have labelled the hydrogen absorption lines. All other prominent absorption features are He I lines.}
\label{fig:Spectrum}
\end{center}
\end{figure}

We then obtained an additional fifty-eight 2-minute exposures using the Low Resolution Imaging Spectrometer (LRIS) \citep{1995PASP..107..375O} on the 10-m W.\ M.\ Keck I Telescope on Mauna Kea, with the 400/8500 grism and a 2x2 binning on the blue arm. Eighteen of these exposures were obtained on Nov 15, 2017 and the remaining forty on March 19th, 2018. These exposures were sufficiently short to allow us to create a time resolved radial velocity curve (Figure \ref{fig:KeckChimera}) and also assemble a co-added spectrum for atmospheric fitting with minimal broadening of the features from the Doppler shifts (Figure \ref{fig:Spectrum}). In individual exposures, we averaged an SNR of 4-5, with a wavelength coverage from 3400-5600 angstroms, and a resolution of approximately $\frac{\lambda}{\Delta\lambda}=700$. In order to ensure wavelength stability, we took a HeNeArCdZn arc at the telescope position of the object. For both nights, we used five consecutive dome flats and five bias frames to perform calibrations, and reduced the data using the lpipe pipeline \citep{2019PASP..131h4503P}.

\section{Analysis and Results}

\subsection{Discovery}
\begin{figure}[htpb]
\begin{center}
\includegraphics[width=0.47\textwidth]{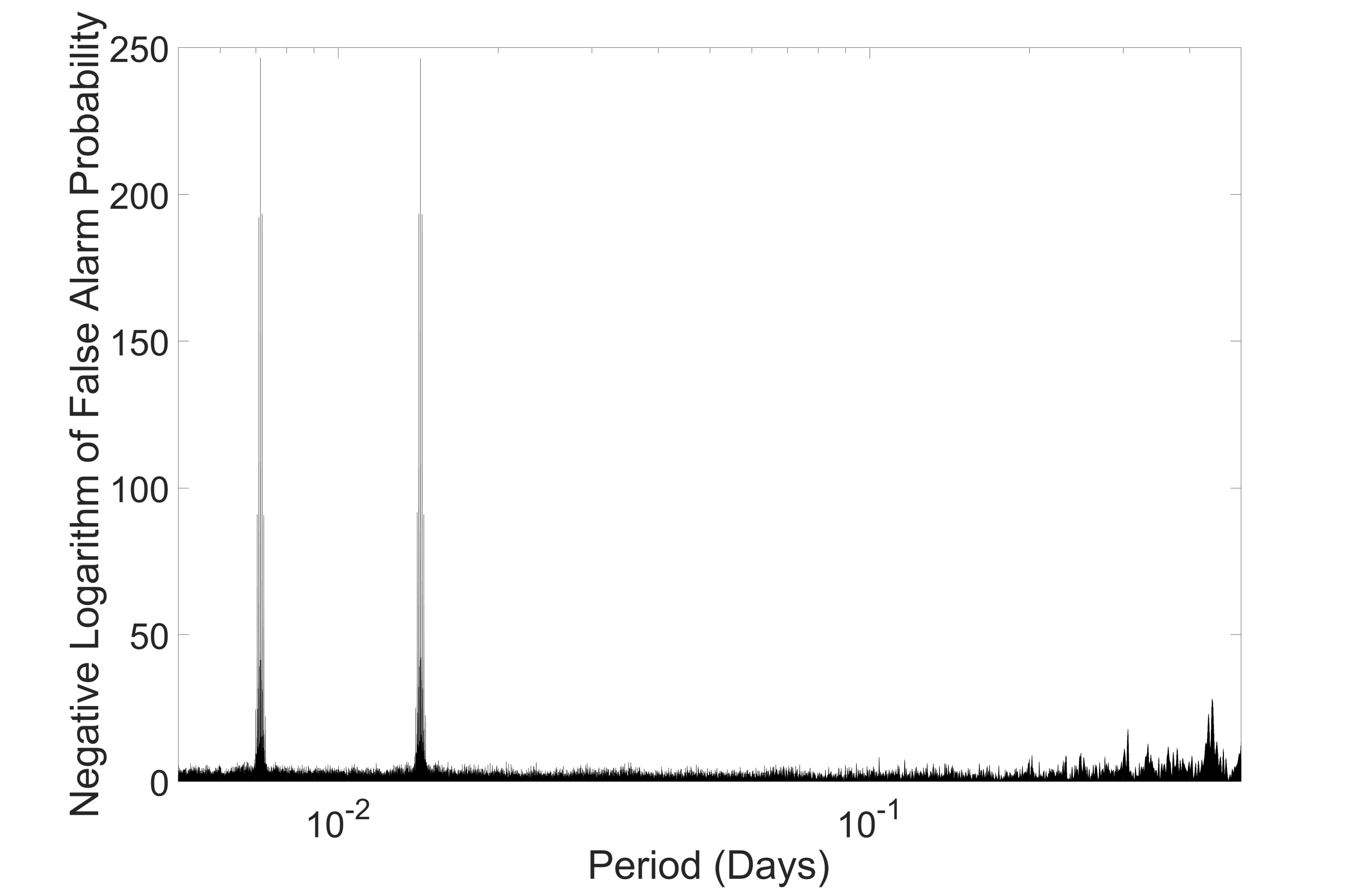}
\caption{A power spectrum of PTF J0533+0209 generated using a multiharmonic analysis of variance routine. The spectrum clearly illustrates two strong signals, one corresponding to half of the orbital period (due to the ellipsoidal modulation), and another strong feature at the orbital period, due to Doppler beaming and gravity darkening. Note that the Y axis is in units of the negative logarithm of the formal false alarm probability reported by the algorithm, which is defined as the probability of a signal with no periodic component at this frequency producing a feature of this amplitude in the power spectrum \citep{2018ApJS..236...16V}. The feature seen at long periods occurs near 0.5 days, and is likely due to the sidereal day.}
\label{fig:PS}
\end{center}
\end{figure}

\begin{figure}[htpb]
\begin{center}
\includegraphics[width=0.47\textwidth]{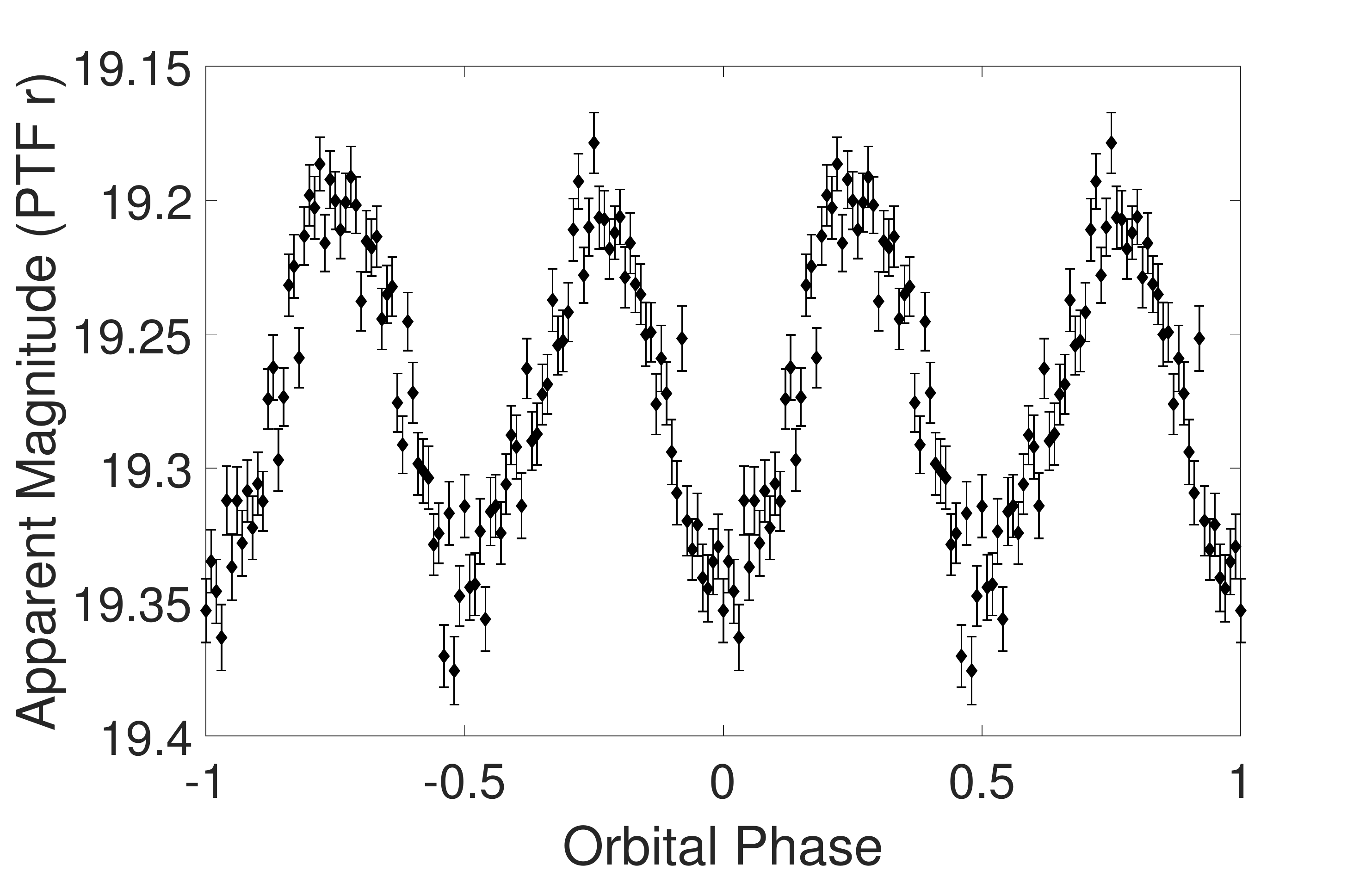}
\caption{The phase folded binned PTF $r$ band lightcurve of PTF J0533+0209, illustrating the strong ellipsoidal modulation which enabled the detection of the source via its optical periodicity.}
\label{fig:PTFLC}
\end{center}
\end{figure}

We discovered the PTF J0533+0209 by applying the multi-harmonic analysis of variance (MHAOV) period finding routine to the sample of color selected lightcurves \citep{1996ApJ...460L.107S}, searching periods from 7.2 minutes to 1000 days. PTF J0533+0209 exhibits strong ellipsoidal modulation in its PTF lightcurve (Figure \ref{fig:PTFLC}), and thus exhibits significant power at half the orbital period in its power spectrum (Figure \ref{fig:PS}). We used MHAOV because of its superior sensitivity to sharp non-sinusoidal periodic features such as eclipses while still remaining sensitive to sinusoidal lightcurves such as that of PTF J0533+0209. We used the implementation of MHAOV available in the Vartools package \citep{2016A&C....17....1H}. In the initial search for periodic objects, we use heliocentric Julian dates as timestamps. However, all timestamps in this publication have been corrected to mid-exposure barycentric Julian dates.

\subsection{Orbital Dynamics}
We measured the Doppler shift of the absorption lines in each time-resolved spectrum by fitting the centroids of the Hydrogen Balmer lines of H$_{\beta}$ and H$_{\gamma}$, as well as the neutral helium lines at wavelengths of 4471, 4713, and 4921 \AA. All of these lines belong to PTF J0533+0209B. We derived the velocity semi-amplitude of PTF J0533+0209B, $K_B$, by performing a weighted least squares fit of a sinusoid to the combined RV data, with a fixed $\omega$ corresponding to the orbital period derived from the PTF data. We derive a velocity semi-amplitude of $K_B=618.7\pm6.9$ km~s$^{-1}$ and a systemic velocity of $\gamma=76.0\pm4.3$km~s$^{-1}$, where we have obtained the 1 sigma error bars from the covariance matrix of the least squared fit to the time resolved radial velocity measurements. The mass function of PTF J0533+0209A is given by:
\begin{equation}
    \label{eq:BMF}
    \frac{M_A^3{\sin}^3(i)}{(M_A+M_B)^2}=\frac{PK_B^3}{2\pi G}
\end{equation}

where $M_A$ is the mass of PTF J0533+0209A, $M_B$ is the mass of PTF J0533+0209B, $i$ is the orbital inclination, and $\rm G$ is the gravitational constant.

Thus, using Equation \ref{eq:BMF}, we can constrain the relationship between the physical parameters of interest in the left hand side of the equation with the measured $P$ and $K_B$.

\subsection{Atmospheric Fitting}

We calculated a grid of DBA (helium-dominated with traces of hydrogen) white dwarf atmosphere models to fit the spectra of PTF J0533+0209B (we found no evidence of any spectroscopic features associated with PTF J0533+0209A). The grid spans the effective temperature range of  $11\,000 \leq T_{\rm eff} \leq 30\,000$ K in steps of $1\,000$ K, the surface gravity range of $5.5 \leq \log{g} \leq 9.0$ in steps of 0.5 dex and the hydrogen-to-helium number density ratio of $-1.0 \leq \log{\rm{H/He}} \leq -10.0$ in steps of 0.5 dex. A detailed description of the model atmosphere code can be found in \cite{bergeron2011}.

Analogous to the well established spectroscopic method used for DA white dwarfs, we compared the absorption line profiles of the continuum-normalised spectrum of PTF J0533+0209B with our newly computed models. In our fit we included a E(B-V) reddening of 0.13 and assumed an extinction constant $A_{\rm V}= 3.1$ \citep{2019arXiv190502734G}.

\begin{figure}[htpb]
\begin{center}
\includegraphics[width=0.47\textwidth]{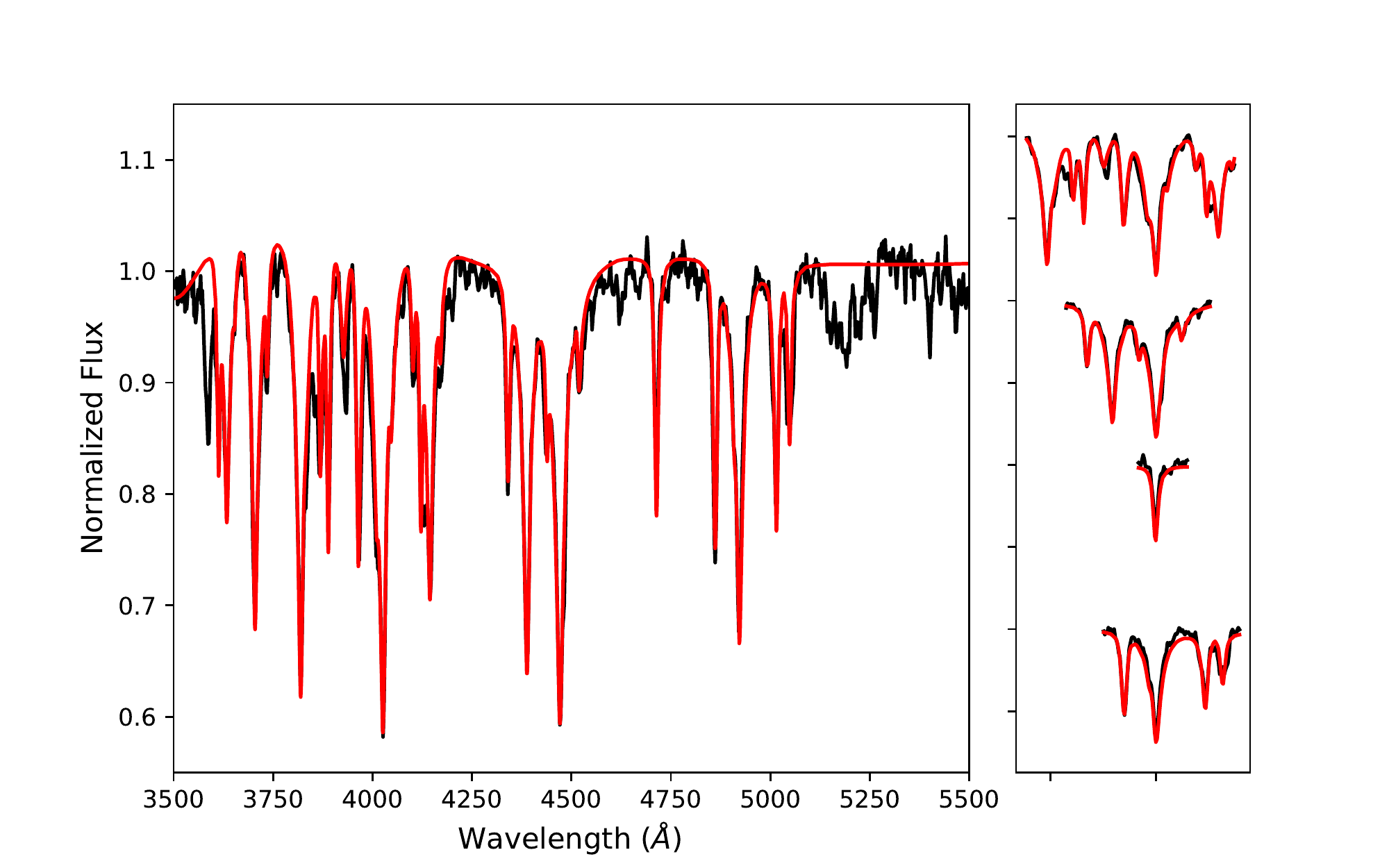}
\caption{The normalized coadded LRIS spectrum of the object is shown in black. In red, we have overlplotted the best fit DBA atmospheric model. The four right hand panels illustrate zoomed in plots of the atmospheric fit to the absorption lines.}
\label{fig:Specfit}
\end{center}
\end{figure}

The best-fit model solution corresponds to $T_{\rm eff}= 20\,000 \pm 800$ K, $\log{g}=6.3 \pm 0.1$, and $\log{\rm{H/He}} = -2.7\pm 0.1$ (Figure \ref{fig:Specfit}). For limb- and gravity-darkening coefficients, an additional grid of atmosphere models was computed for $\log{\rm{H/He}} = -2.56 $,  $19\,250\leq T_{\rm eff} \leq 20\,500$ K in steps of 250K and $5.4 \leq \log{g} \leq 7.2$ in steps of 0.6 dex. For each model we calculated the specific intensity at 20 different angles.

\subsection{Lightcurve Modelling}

Because this binary exhibits significant ellipsoidal modulation, we can impose constraints on the system parameters by modelling its lightcurve. We see ellipsoidal variations in the lightcurve because PTF J0533+0209B is being tidally deformed, and over the course of the orbit, we see different geometric cross sections of the object. Equation \ref{eq:Ellipsoidal} gives the fractional flux variation due to ellipsoidal modulation:

\begin{equation} \label{eq:Ellipsoidal}
\frac{\Delta F_{\rm ellipsoidal}}{F}=0.15\frac{(15+u)(1+\tau)}{3-u}\left(\frac{R}{a}\right)^3q\sin^2{i}
\end{equation}

\noindent where the fractional flux variation due to ellipsoidal variations, $\frac{\Delta F_{\rm ellipsoidal}}{F}$ depends on the ratio of the radius of the tidally deformed object to the semi-major axis, $R/a$, the tidally deformed component's linear limb darkening coefficient $u$, its gravity darkening coefficient $\tau$, the mass ratio of the binary $q$, and the inclination $i$ \citep{2012MNRAS.422.2600B}.

Because this object is strongly deformed, we must also account for an additional effect known as gravity darkening, which visibly manifests itself as an asymmetry in the two flux minima of the ellipsoidal modulation. This is a result of the flux emitted on the surface of the deformed object near the first Langrange point (L1) being less than that emitted from the surface near the L2 point. Von Zeipel's theorem \citep{1924MNRAS..84..665V} states that the flux emitted off of the surface element of a star is proportional to the local gravitational field. Although the material at the surface of a tidally deformed star falls on a gravitational equipotential surface $\phi=\rm const$, the gradient of the gravitational potential along this surface is not necessarily constant or symmetric, i.e., $-\nabla \phi=g\neq \rm const$, and this is manifested by an asymmetry between the flux emitted by surface elements at L1 and L2.

We used the ellc package to model the lightcurve \citep{2016A&A...591A.111M}. We modelled both the \textbf{$g^\prime$} and $i^\prime$ lightcurves, adopting a linear limb darkening model, and fixing the linear limb darkening coefficient of the secondary $u_2$ to a value of $0.25$ for the $g^\prime$ lightcurve, and to $0.18$ for the $i^\prime$ lightcurve. We also fixed the passband dependent gravity darkening coefficient of the secondary, $\tau_2$, to 0.25 for the $g^\prime$ lightcurve, and 0.19 for the $i^\prime$ lightcurve. These coefficients were computed using the procedure outlined in \cite{2017A&A...600A..30C}, adjusted for DBA atmospheres (Claret, Cukanovaite \& Burdge 2019, in preparation). We computed and fixed the Doppler beaming factors of PTF J0533+0209B to 1.94 for $g^\prime$ and 1.54 for $i^\prime$ \citep{2003ApJ...588L.117L} using the atmospherically derived temperature of $T_{\rm eff}=20000 \, \rm K$. The free parameters in the model were the ratio of the radii to the semi-major axis for the two components, $R_A/a$ and $R_B/a$, the inclination $i$, the time of minimum light $t_0$, the semi-major axis $a$, and the mass ratio $q=M_B/M_A$. Although we see no sign of luminosity from PTF J0533+0209A in the spectrum or the spectral energy distribution, as an initial test, we conducted an iteration of modelling with the surface brightness ratio, $J$, and albedo of PTF J0533+0209B, $\rm heat_B$, as free parameters to investigate whether PTF J0533+0209B might be irradiated by its unseen companion. We found the solution did not converge to any particular value of these parameters for the $g^\prime$ or $i^\prime$ lightcurves, and that the parameters exhibited no co-variance with any other parameters. We performed the same exercise considering the possibility of PTF J0533+0209A being irradiated by PTF J0533+0209B, and found a similar result. Based on this, for the final iteration of modelling, we assumed no luminosity contribution from PTF J0533+0209A, and omitted accounting for any kind of irradiation effect. The faint GALEX NUV apparent magnitude (see Table \ref{tab:ObservableParms}) also suggests it is unlikely that there is an unseen hot companion.

\subsection{Orbital Period Decay}

Because of the short orbital period and non-interacting nature of this system, we expect a possible measurable orbital decay due to the emission of gravitational radiation \citep{1979Natur.277..437T}. One can determine the orbital frequency derivative, $\dot{f}$, by measuring a change in the phase of the lightcurve over time, $\Delta t_{\rm ellipsoidal}$, as illustrated in Equation \ref{eq:OC}:
\begin{equation} \label{eq:OC}
\Delta t_{ellipsoidal}(t-t_0)= \Big(\frac{1}{2}\dot{f}(t_0)(t-t_0)^2+... \Big) P(t_0)
\end{equation}
where $t_0$ is the reference epoch, $t-t_0$ is the time since the reference epoch, $f(t_0), \dot{f}(t_0),$ $ \rm etc,$ are the orbital frequency and its derivatives at the reference epoch, and $P(t_0)=\frac{1}{f(t_0)}$ is the orbital period at the reference epoch.

\begin{figure}[htpb]
\begin{center}
\includegraphics[width=0.5\textwidth]{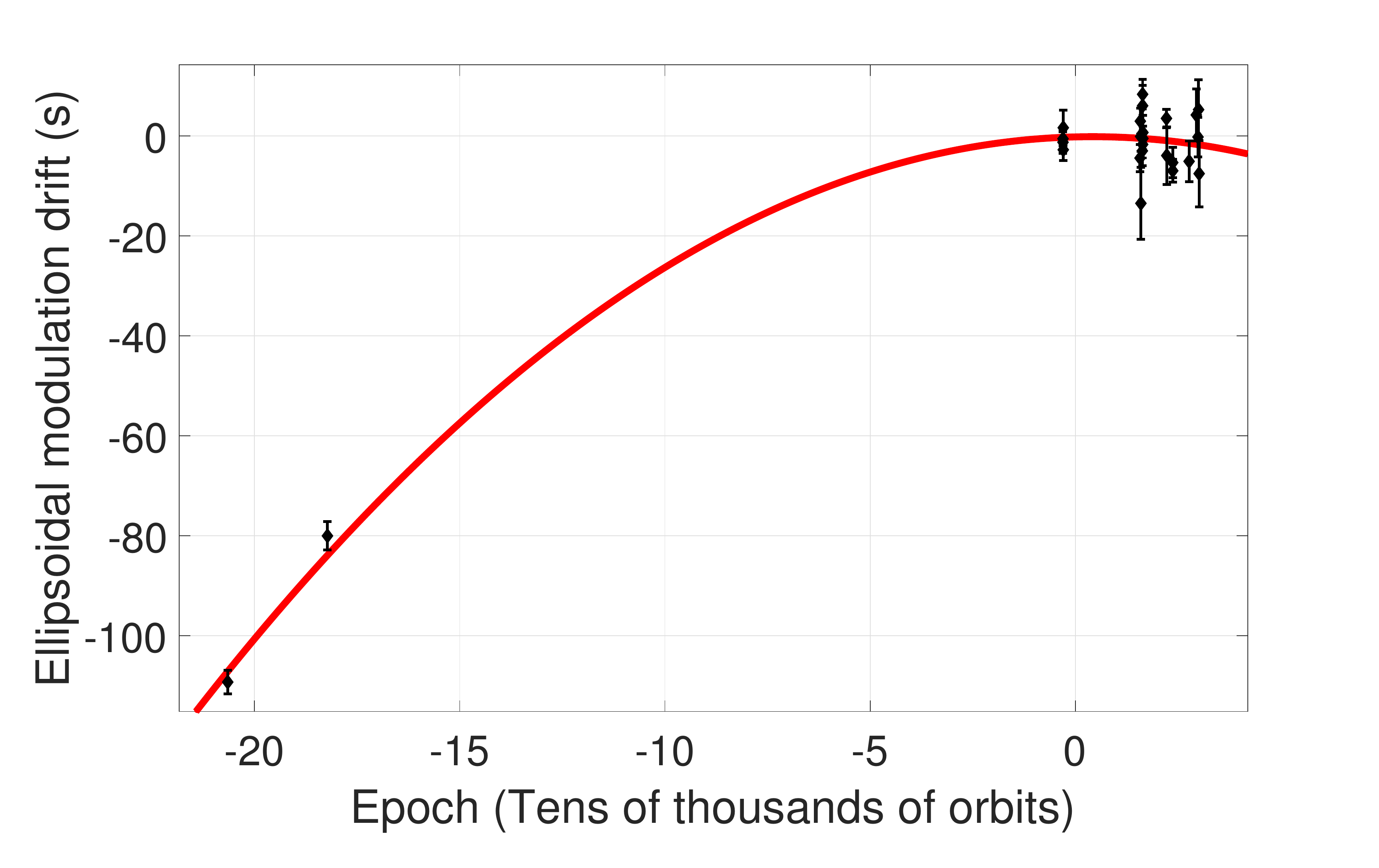}
\caption{The measured drift of the ellipsoidal modulation of PTF J0533+0209 over time. The two initial points, both originating from Palomar Transient Factory data acquired nearly a decade before the other epochs, illustrates that this signal has shifted in phase by over $100\rm s$ since these observations, consistent with the expectations of orbital decay via gravitational radiation.}
\label{fig:Decay}
\end{center}
\end{figure}

In order to extract the timing epochs from PTF, KPED, and CHIMERA data, we used a least squares fit of a sinusoid to the data and measured the phase of this sinusoid. We then used a least squares fit of a quadratic to these time stamps as a function of epoch (Figure \ref{fig:Decay}) to measure the orbital decay rate, $\dot{P}$. We report the derived ephemeris in Table \ref{tab:Parameters}.

If purely due to general relativity, the decay should be governed by the expression given in Equation \ref{eq:Decay}:
\begin{equation}\label{eq:Decay}
   \dot{f}_{GW}=\frac{96}{5}\pi^{\frac{8}{3}} \left(\frac{GM_c}{c^3}\right)^\frac{5}{3}f_{GW}^\frac{11}{3}
\end{equation}
where $M_c=\frac{(M_1M_2)^{\frac{3}{5}}}{(M_1+M_2)^{\frac{1}{5}}}$ is the chirp mass, $c$ is the speed of light, and $f_{GW}=\frac{2}{P}$ is the gravitational wave frequency. Thus, we can use the measured decay rate to constrain the chirp mass of the system. Such decay has been observed in several known detached double degenerate systems, including ZTF J1539+5027 \citep{Burdge2019} and SDSS J0651+2824 \citep{2012ApJ...757L..21H}, as well as some accreting systems such as HM Cancri \citep{2005ApJ...627..920S}.

\subsection{Tidal Contribution to Orbital Decay}

Equation \ref{eq:pdottide2} gives an estimate for tidal contribution to the measured orbital decay:
\begin{equation}
    \label{eq:pdottide2}
    \frac{\dot{P}_{\rm tide}}{\dot{P}_{\rm GW}} \simeq \frac{ 3 (M_{A}+M_{B})}{M_A M_B} \bigg[ \kappa_A M_A \bigg( \frac{R_A}{a}\bigg)^2 + \kappa_A M_2 \bigg( \frac{R_B}{a}\bigg)^2 \bigg]
\end{equation}
where $\dot{P}_{\rm tide}$ is the tidal contribution to the orbital decay, expressed as a fraction of the contribution to orbital decay from the emission of gravitational radiation, $\dot{P}_{\rm GW}$, which is a function of the ratio of the radii to the semi-major axis of the two components, the component masses, and also $\kappa_a$ and $\kappa_b$, dimensionless constants reflecting the internal structure of the white dwarf \citep{Burdge2019}.

Using the masses derived from the combined analysis, as well as $\kappa_a=0.14$ and $\kappa_b=0.066$ based on white dwarf models, we estimate a tidal contribution of $\frac{\dot{P}_{\rm tide}}{\dot{P}_{\rm GW}}\simeq0.02$. We correct the chirp mass inferred from $\dot{P}$ by accounting for this tidal contribution, though it does not significantly alter the solution, as it is less than a sixth of the measurement uncertainty on $\dot{P}$. As the measurement of $\dot{P}$ improves with time, the tidal contribution will become large compared to the measurement uncertainty, and will become the dominant source of uncertainty on the chirp mass.

\subsection{Parameter Estimation}

\begin{figure}[htpb]
\begin{center}
\includegraphics[width=0.47\textwidth]{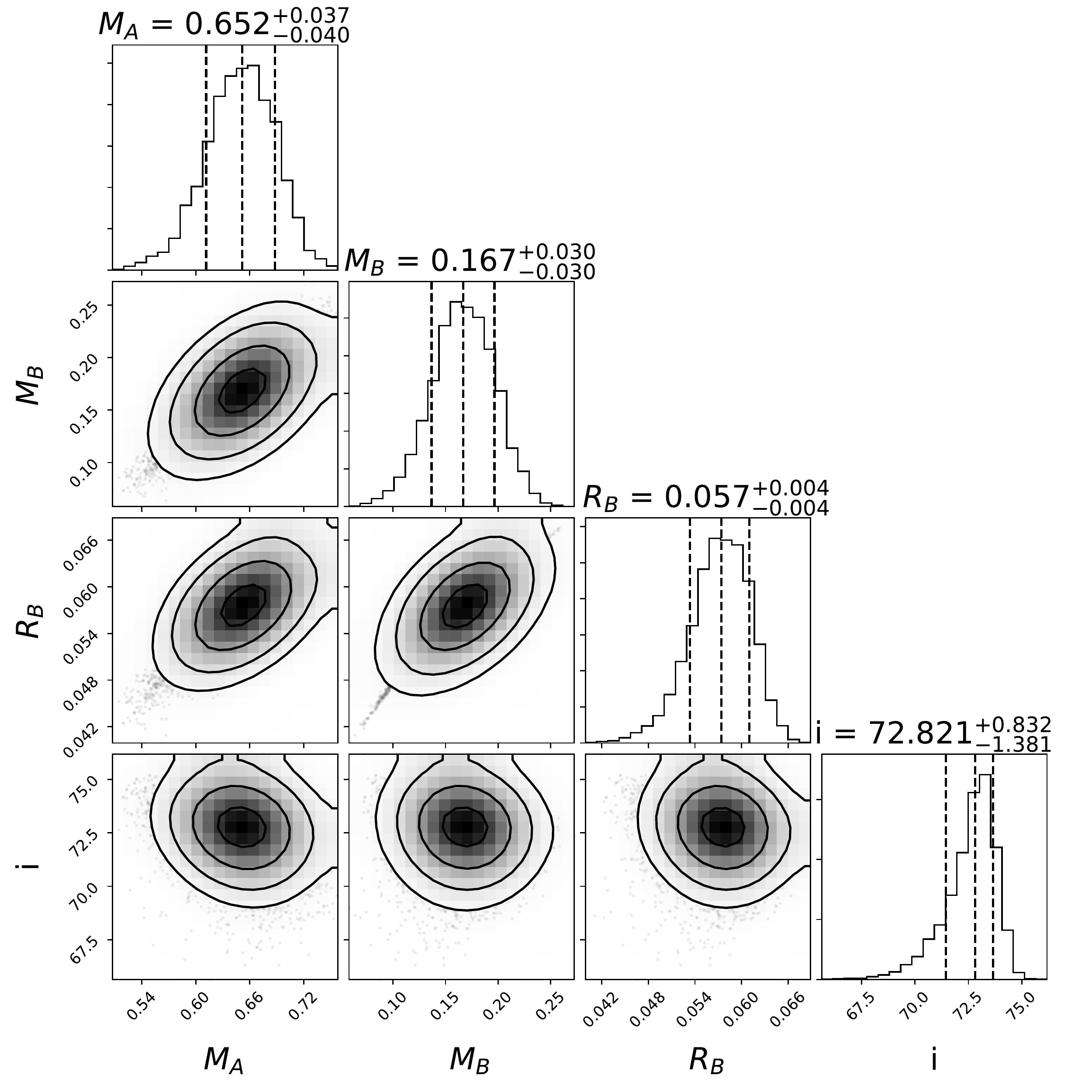}
\caption{Corner plots of the physical parameters inferred as a result of the analysis combining the lightcurve modelling with radial velocity and chirp mass constraints.}
\label{fig:Corner}
\end{center}
\end{figure}

In order to estimate the physical parameters of the system, we combined the measurements of orbital kinematics and orbital period decay with the lightcurve models fit to the CHIMERA $g^\prime$ and $i^\prime$ data taken on December 14-15 of 2017. The atmospheric fit parameters only entered this modelling as a basis for deriving the limb and gravity darkening coefficients used in the lightcurve modelling. The free parameters of this analysis were the masses $M_A$ and $M_B$, the inclination $i$, the radius $R_B$, and the time of minimum light $t_0$. We used the multinest algorithm \citep{2009MNRAS.398.1601F} to sample over these free parameters, and present the corner plots of the resulting analysis in Figure \ref{fig:Corner}. The inferred parameters from this analysis, as well as the atmospheric analysis and timing analysis, are listed in Table \ref{tab:Parameters}. 

\begin{table}[h!]
\renewcommand{\thetable}{\arabic{table}}
\centering
\caption{Table of physical parameters} \label{tab:Parameters}
\begin{tabular}{cD@{$\pm$}D}
\tablewidth{0pt}
\hline
\hline
   $\rm M_A$ & $0.652^{+0.037}_{-0.040}\,M_\odot$   \\
   \hline
   $\rm M_B$ & $0.167^{+0.030}_{-0.030}\,M_\odot$     \\
   \hline
   $\rm i$ & $72.8^{+0.8}_{-1.4}\,\rm degrees $  \\
   \hline
   $\rm R_B$ & $0.057^{+0.004}_{-0.004}\,R_\odot$  \\
   \hline
   $\rm T_B$ & $20000^{+800}_{-800}\,K$  \\
   \hline
   $\rm log(g)_{B}$ & $6.3^{+0.1}_{-0.1}$  \\
   \hline
   $\log{\rm{H/He}}_{\rm B}$ & $-2.7^{+0.1}_{-0.1}$  \\
   \hline
   $T_{0}$ & $2458145.096848^{+0.000046}_{-0.000046}\, \rm BJD_{TDB}$  \\
   \hline
   $P(T_{0})$ & $1233.97298^{+0.00017}_{-0.00017}\,\rm s$  \\
   \hline
   $\dot{P}(T_{0})$ & $(-3.94\pm0.80)\times10^{-12}\,\rm s\,\rm s^{-1}$  \\
  \hline
  \hline

\end{tabular}
\end{table}

\subsection{Distance Estimate}

PTF J0533+0209 has a corresponding entry in the Gaia DR2 catalog \citep{2018A&A...616A...1G}, with a measured parallax of $\bar{\omega}=0.4741$ and uncertainty $\sigma_{\bar{\omega}}=0.4786$. Since we are in a regime where $\sigma_{\bar{\omega}}>\bar{\omega}$, we cannot simply infer the distance as the reciprocal of the parallax. We adopt an exponentially decreasing space density prior \citep{2018AJ....156...58B}, with an assumed characteristic length scale of $400\, \rm pc$ \citep{2018MNRAS.480..302K}. Using this technique, we infer a distance of $D=1.5^{+0.7}_{-0.5}\, \rm kpc$, and we adopt this distance for our estimate of the \emph{LISA} gravitational wave strain of the system. If we assume a length scale of $200\, \rm pc$, the solution becomes $D=1.05^{+0.4}_{-0.3}\, \rm kpc$, and for a length scale of $800\, \rm pc$, $D=2.3^{+1.4}_{-0.9}\, \rm kpc$. We would like to note that using our inferred radius and temperature, we estimate a distance of $D=2.7^{+0.2}_{-0.2}\, \rm kpc$, farther than any of these estimates. 

\subsection{Gravitational Wave Strain}

To calculate the characteristic strain \citep{2017MNRAS.470.1894K}, $S_c$, we use the expression in Equation \ref{eq:strain}, where $c$ is the speed of light, and $T_{obs}$ is the operation time of the \emph{LISA} mission.
\begin{equation}
    \label{eq:strain}
   S_{c}=\frac{2(GM_{c})^{5/3}(\pi f)^{2/3}}{c^{4}D}\sqrt{fT_{obs}}
\end{equation}
When computing the characteristic strain, we marginalized over the posterior distributions for the masses derived from our combined analysis, and the distance estimated using the parallax measurement and the assumption of a $400 \, \rm pc$ length scale. The point and uncertainties corresponding to PTF J0533+0209 in Figure \ref{fig:LISA} reflects the posterior distribution of the characteristic strain of this object.

However, in order to compute the estimated SNR of the source in \emph{LISA}, we must account for instrument response, as well as the source position in the sky, and the source inclination. Thus, to compute the SNR, we calculate the amplitude of the signal at the detector, $A=\sqrt{|F_{+}|^{2}|h_{+}|^{2}+|F_{\times}|^{2}|h_{\times}|^{2}}$, which depends on the two gravitational wave polarizations, $h_{+}$ and $h_{\times}$, and \emph{LISA}'s response patterns in coupling to these polarizations, $F_{+}$ and $F_{\times}$ \citep{2017MNRAS.470.1894K}.

\section{Discussion}

\subsection{LISA gravitational wave source}

\begin{figure}[htpb]
\begin{center}
\includegraphics[width=0.47\textwidth]{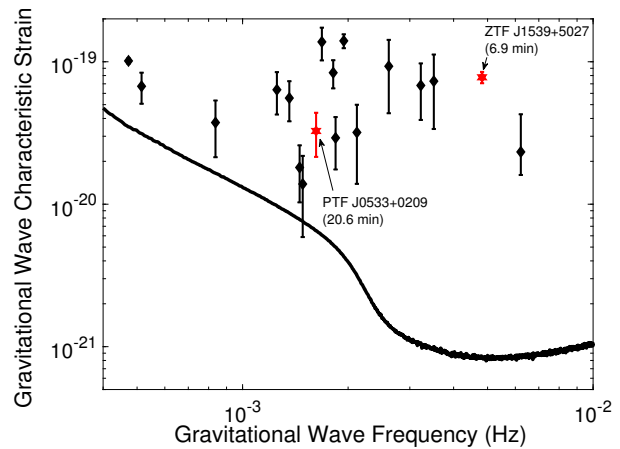}
\caption{A plot of the characteristic gravitational wave strains of the known \emph{LISA} detectable binaries after 4 years of integration and the \emph{LISA} sensitivity curve (shown as the smooth black curve). The black diamonds are sources reported in \cite{2018MNRAS.480..302K}, whose errors have been estimated using Gaia parallaxes, with the exception of HM Cancri, which we assigned a uniform prior in distance of 4.2-20 kpc. The red stars indicate the two \emph{LISA} gravitational wave sources discovered in the recent survey for \emph{LISA} sources using PTF/ZTF \citep{Burdge2019}.}
\label{fig:LISA}
\end{center}
\end{figure}

Based on its component masses, inclination, and distance of $D=1.5^{+0.7}_{-0.5}\, \rm kpc$ based on the parallax measurement and $400 \, \rm pc$ length scale, we estimate that \emph{LISA} will detect PTF J0533+0209 with a signal to noise ratio of $8.4^{+4.2}_{-3.0}$ at the end of the nominal four year mission lifetime (Figure \ref{fig:LISA}). If we instead use the distance of $D=2.7^{+0.2}_{-0.2}\, \rm kpc$ inferred from the radius, temperature, and apparent magnitude of the object, the estimated signal to noise after four years is $4.7^{+0.9}_{-0.9}$. By the launch of \emph{LISA} in 2034, we estimate that the orbital decay of PTF J0533+0209 will have caused the ellipsoidal modulation to have drifted by $>900s$ in phase since the initial PTF epochs, allowing for a precise constraint on the chirp mass with regular monitoring over this time period. With the chirp mass constrained, the strain measured by \emph{LISA} will enable a direct probe of the distance to the object, helping verify the estimates of radius and surface temperature in this analysis . Perhaps the best example of \emph{LISA}'s utility is that it will be able to provide an independent measurement of the inclination angle of the system \citep{2012A&A...544A.153S}. This is because the gravitational wave signal can be decomposed into two polarization components, $h_{+}$ and $h_{\times}$, and the strain amplitudes of these components include factors of $\big(1+\cos^2{(i)} \big)$ and $2\cos{(i)}$, respectively \citep{1987thyg.book..330T}. Combined with the measurements of $K_B$, and $\dot{P}$, this will provide a constraint on both component masses independent of the model-dependent fit of the ellipsoidal modulation.

\subsection{Binary Formation Models}

\begin{figure}[htpb]
\begin{center}
\includegraphics[width=0.47\textwidth]{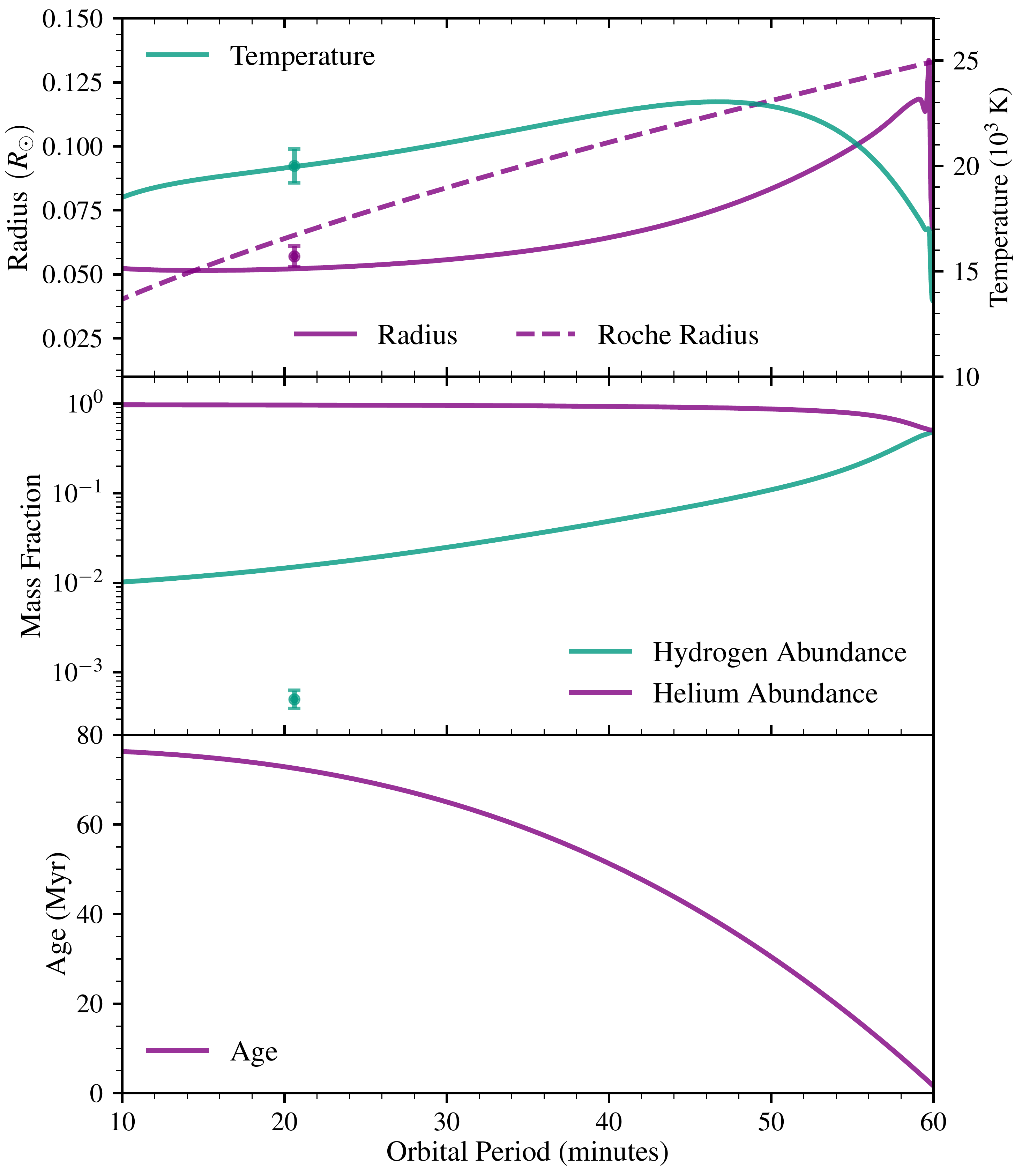}
\caption{Binary Evolution Model of PTF J0533+0209, beginning with a $0.19 \, M_\odot$ ELM WD and a $0.66 \, M_\odot$ CO WD in a one-hour orbit. The ELM WD cools and contracts, and at an orbital period of 20 minutes it has nearly the same surface temperature and radius as the ELM WD in PTF J0533+0209, whose properties are denoted by colored points with error bars. The surface hydrogen abundance decreases due to extra mixing in the model (see text), though it remains larger than our spectroscopic measurements. }
\label{fig:20min}
\end{center}
\end{figure}  

PTF J0533+0209 provides interesting constraints on compact binary formation physics. In particular, the combination of the extremely low-mass white dwarf (ELM WD), high surface temperature, and low surface hydrogen (H) abundance is difficult to understand, since most ELM WDs are H-rich (DA) WDs due to helium (He) sedimentation. H-deficient CO WDs, such as DB WDs, are thought to originate from AGB stars undergoing a late He-burning shell flash during the early WD cooling phase. The shell flash induces convection that subsumes the H-envelope, mixing H deep into the star where it is quickly burnt \citep{althaus:05}. However, the He-core ELM WD in PTF J0533+0209 could not have undergone such a He-burning pulse, so the H could not be destroyed in the same way that it is for DB WDs. Nor is it likely the H has been totally stripped due to mass transfer like in AM CVn systems, because there is no reason to think that mass transfer has yet begun in this system. Several H-poor proto-ELM WDs have been discovered (see e.g., \citealt{kaplan:13,gianninas:16}), though these have higher H-abundances and low surface gravities, and can likely be explained by weaker gravitational settling opposed by rotational mixing \cite{istrate:16}.

To understand the formation of PTF J0533+0209, we constructed models of binary stars using the MESA stellar evolution code \citep{paxton:11,paxton:13,paxton:15}. First, we evolve a $1.2 \, M_\odot$ star up the red giant branch until its core helium mass is $0.187 \, M_\odot$. We then strip the outer layers until the star has a mass of $0.19 \, M_\odot$, leaving it with $10^{-3} \, M_\odot$ of hydrogen, and place it in a binary with a $0.66 \, M_\odot$ companion in a 1 hour orbit. This mimics a common envelope (CE) event that births a compact binary system with masses consistent with our measurements for PTF J0533+0209. We evolve this system using MESA's binary module, allowing for orbital decay via gravitational radiation, while simultaneously evolving the structure of the ELM WD. We enforce tidal synchronization at all times, as short-period binary WDs are expected to be nearly tidally synchronized \citep{fullerwd:12}. Rotational mixing is included via MESA's implementation of Eddington-Sweet circulation with  \verb|am_D_mix_factor = 1| and \verb|D_ES_factor = 1|. We note that this amount of rotational mixing is very optimistic, as it is a factor of 30 larger than expected \citep{heger:00}, yet we shall see below that even this excessive estimate of rotational mixing cannot explain the observed hydrogen depletion.

Figure \ref{fig:20min} shows the evolution of the system described above. As the system ages, the WD contracts, and the  model radius is $R \simeq 0.05 \, R_\odot$ at a 20 minute orbital period. Because of extra mixing in our model (see below), the temperature initially increases as H is mixed into the interior and burned, but the temperature then decreases as H is depleted. The mass, temperature, and radius of our model at 20 minutes are all approximately consistent with PTF J0533+0209. If the post-CE period is substantially longer than one hour, the ELM WD cools to temperatures lower than observed. So, the system can only be explained by the CE formation channel if the ELM WD is born at a fairly short orbital period. 

While the match with observations above is encouraging, it is difficult to explain the low H abundance from this model, which initially has a roughly equal surface mass fraction of H and He after the CE event. The Eddington-Sweet mixing keeps the outermost layers of the WD well-mixed due to the tidally enforced rapid rotation of the ELM WD, but it does not mix H into the deep interior. In the limited number of models we have explored, rotational mixing is not enough to mix H down into the He core where it can be burned, allowing the surface H to be depleted. In order to greatly deplete the surface H abundance, we have added an ad-hoc mixing diffusivity of $D_{\rm mix} = 2.5 \times 10^3 \, {\rm cm}^2/{\rm s}$, which causes H to mix deep enough to burn. Figure \ref{fig:20min} shows that the surface mass fraction of H is depleted by a factor of 50 down to $\sim \! 10^{-2}$ in this model, though it is still much larger than our measurements indicate. Without the extra mixing, the surface mass fraction remains nearly constant. Our models do not include gravitational settling and diffusion, which would cause the surface mass fraction of H to {\it increase} as He gravitationally settles. The burning of H in our model keeps it warmer compared to a model without mixing, so our model cools slower than a normal WD of the same mass.

This model is not meant to be a ``fit" to the observed properties, and some other combination of mass, initial orbital period, extra mixing, etc., can likely provide a better match to the data. Our main claim is that some sort of extra mixing or other method of removing H from the ELM WD must be at play. Since PTF J0533+0209 is in a very short-period binary compared to other H-rich ELMs, we suspect the extra mixing may be related to rotational or tidal effects that are much stronger in tight binaries. Also, if PTF J0533+0209 evolved from the CE channel, the system was likely born at an orbital period less than a couple hours in order for the ELM WD to be so hot. Based on the findings of \cite{fuller2013dynamical}, we find that tidal heating is unlikely to heat up the ELM WD to the observed temperature at an orbital period of 20 minutes. From equation 10 of \cite{Burdge2019}, the ``tidal" temperature to which the ELM WD can be heated is only $\sim \! 7000 \, {\rm K}$, much below the observed temperature. A tidally induced nova \citep{fullernova:12} could heat up the ELM WD, but these events are not expected to occur at such long orbital periods. 

It may be possible that H-burning flashes contribute to the depletion of H near the surface, as such flashes in a compact binary will cause the ELM to overflow its Roche lobe and lose much of its expanding H envelope. The model described above had a small enough initial H mass that it did not encounter such flashes, but we have attempted to simulate flashing models with a larger initial H mass. For numerical reasons, we have not been able to simulate the flash-induced mass loss in a compact binary, however, the models indicate that the H at the base of the envelope would not overflow the Roche lobe during these events and would likely be retained. Hence, we believe H flashes may have contributed to the depletion of H, but some extra mixing  (e.g., rotational mixing, or mixing induced via tidally excited gravity waves) is likely required to further deplete the remaining H to the observed level. One possibility is that the H-burning flashes induce more mixing below the convective H-burning shell than predicted by our models, a phenomenon also thought to occur during classical novae (e.g., \citealt{alexakis:04,denissenkov:13}). This could mix H downward where it is burnt and dilute the H shell with dredged up He, potentially reducing the observed H-abundance to the level observed.

We have also attempted to create a system like PTF J0533+0209 from a stable mass transfer scenario like that described in \cite{sun:18,tauris:18,li:19}. We evolve systems near the orbital period  bifurcation \citep{podsiadlowski:03} such that the donor is evolved (and hence H-poor) and the binary shrinks to sub-hour orbital periods via magnetic braking and gravitational radiation. This channel typically forms low-mass ELM WDs with masses in the range $0.15 \, M_\odot \lesssim M \lesssim 0.2 \, M_\odot$ that detach at very short orbital periods, so it can certainly produce systems similar to PTF J0533+0209. However, the donor in these models usually lower surface temperature and higher H abundance than observed in PTF J0533+0209. Hence, while it might be possible to form a system like PTF J0533+0209 through stable mass transfer, we consider the CE channel to be more likely. 

\subsection{Evolutionary fate}

Based on its measured orbital decay rate, PTF J0533+0209 has a characteristic decay timescale of  $\tau_c=\frac{3}{8}\frac{P}{|\dot{P}|}\approx 3,700,000$\,years. However, in a fraction of this time, PTF J0533+0209B will overflow its Roche lobe and initiate mass transfer onto PTF J0533+0209A, which occurs at an orbital period of $\approx \! 14$ minutes in Figure \ref{fig:20min}, though precisely when this occurs depends on the rate at which PTF J0533+0209B cools and shrinks compared to its merger timescale. The binary will continue to evolve to shorter orbital periods due to the emission of gravitational radiation and eventually mass transfer will begin to remove mass from the degenerate helium core of PTF J0533+0209B \citep{2012ApJ...758...64K}, which will expand in response to this mass loss, consequentially increasing the mass transfer rate and producing a feedback loop. Given the mass ratio of the binary, this means that PTF J0533+0209 will likely evolve into an AM CVn system at longer orbital periods \citep{2004MNRAS.350..113M}, or alternatively, could result in a merger producing an R Coronae Borealis type star \citep{1971AcA....21....1P}.

\section{Conclusion}

We have discovered PTF J0533+0209, a detached double white dwarf binary with a 20 minute orbital period using purely its ellipsoidal modulation. This more luminous component of this system exhibits a DBA atmosphere, unlike most of its double white dwarf counterparts, which is not easy to explain from standard binary formation and stellar evolution models. We have detected orbital decay in this system using archival time domain data, and estimated that the gravitational waves which the system emits should be detectable by \emph{LISA}.

Modern synoptic time domain surveys are not only becoming wider and deeper, but crucially, increasingly well sampled as a consequence of their large fields of view and roboticized observing capabilities. Thus, before \emph{LISA} begins to decode the gravitational-wave signal from tens of thousands of galactic binaries in the Milky Way, these surveys hold the key to uncovering a large population of these systems. Some of these binaries, like ZTF J1539+5027 \citep{Burdge2019}, will be extremely well calibrated via their optical behavior because of eclipses and/or a double-lined spectroscopic nature. However, systems like PTF J0533+0209, which exhibit a handful of very well defined observables like a single radial velocity amplitude, ellipsoidal modulation, orbital decay rate, etc., will be far more numerous than their eclipsing, double-lined counterparts. Objects like PTF J0533+0209 demonstrate the scientific promise of the \emph{LISA} era of astrophysics because \emph{LISA} will transform such systems from being constrained by highly model dependent assumptions to over-constrained systems which will instead test the fidelity of physical models.

\section*{Acknowledgments}

K.B.B thanks the National Aeronautics and Space Administration and the Heising Simons Foundation for supporting his research. JF acknowledges support from an Innovator Grant from The Rose Hills Foundation and the Sloan Foundation through grant FG-2018-10515.
 
Based on observations obtained with the Samuel Oschin Telescope at the Palomar Observatory as part of the Palomar Transient Factory project, a scientific collaboration between the California Institute of Technology, Columbia University, Las Cumbres Observatory, the Lawrence Berkeley National Laboratory, the National Energy Research Scientific Computing Center, the University of Oxford, and the Weizmann Institute of Science.
 
The KPED team thanks the National Science Foundation and the National Optical Astronomical Observatory for making the Kitt Peak 2.1-m telescope available. The KPED team thanks the National Science Foundation, the National Optical Astronomical Observatory and the Murty family for support in the building and operation of KPED. In addition, they thank the CHIMERA project for use of the Electron Multiplying CCD (EMCCD).

Some of the data presented herein were obtained at the W.M. Keck Observatory, which is operated as a scientific partnership among the California Institute of Technology, the University of California and the National Aeronautics and Space Administration. The Observatory was made possible by the generous financial support of the W.M. Keck Foundation. The authors wish to recognize and acknowledge the very significant cultural role and reverence that the summit of Mauna Kea has always had within the indigenous Hawaiian community. We are most fortunate to have the opportunity to conduct observations from this mountain.

The research leading to these results has received funding from the European Research Council under the European Union’s Horizon 2020 research and innovation programme n.677706 (WD3D)
 
This research benefited from interactions at the ZTF Theory Network Meeting that were funded by the Gordon and Betty Moore Foundation through Grant GBMF5076 and support from the National Science Foundation through PHY-1748958. 

\facility{PO:1.2m (PTF), Keck:I (LRIS), Pan-STARRS, Hale (Chimera, DBSP), KPNO/NOAO (KPED)}



\end{document}